\documentclass[pre,superscriptaddress,twocolumn,a4paper,showpacs]{revtex4-1}
\usepackage{amssymb}
\usepackage{amsmath}
\usepackage{wasysym}
\usepackage{graphicx}
\usepackage{epsfig}
\usepackage{subfigure}
\usepackage{float}
\begin{document}

\title{Individual decision making in  task-oriented groups}

\author{Sandro M. Reia}
\affiliation{Instituto de F\'{\i}sica de S\~ao Carlos,
  Universidade de S\~ao Paulo,
  Caixa Postal 369, 13560-970 S\~ao Carlos, S\~ao Paulo, Brazil}

\author{Paulo F. Gomes}
\affiliation{Instituto de F\'{\i}sica de S\~ao Carlos,
  Universidade de S\~ao Paulo,
  Caixa Postal 369, 13560-970 S\~ao Carlos, S\~ao Paulo, Brazil}
\affiliation{Instituto de Ci\^encias Exatas e Tecnol\'ogicas, Universidade Federal de Goi\'as, 
75801-615  Jata\'{\i}, Goi\'as, Brazil }

\author{Jos\'e F.  Fontanari}
\affiliation{Instituto de F\'{\i}sica de S\~ao Carlos,
  Universidade de S\~ao Paulo,
  Caixa Postal 369, 13560-970 S\~ao Carlos, S\~ao Paulo, Brazil}

\begin{abstract}
The strategies adopted  by   individuals to select  relevant information to pass on are  central to 
understanding  problem solving by groups.  Here we use agent-based simulations to revisit a cooperative  problem-solving scenario where the task is to find the common card  in  decks distributed to the group members.  The agents can display only a sample of their cards and we explore different strategies to select those samples based on the confidences assigned to the cards. An agent's confidence that a particular card is the correct one is given by  the number of times it observed that card in the decks of the  other agents.
We use a Gibbs distribution to select the card samples with  the temperature measuring the strength of a noise that prevents the agents to correctly rank the cards.  The group is guaranteed to find the common card in all runs solely  in the infinite temperature limit, where the cards are sampled regardless of their confidences. In this case, we  obtain the  scaling form of the  time constant
that characterizes  the asymptotic exponential decay  of the failure probability. For finite time, however, a finite temperature yields a probability of failure  that is several orders of magnitude lower than in the  infinite temperature limit.  The available experimental results are  consistent  with the decision-making model  for finite temperature only.
\end{abstract}

\maketitle

\section{Introduction}\label{sec:intro}

A central goal of the study of psychologically-inspired   problem-solving systems is the  understanding and modeling of  the human mental processes of decision making  \cite{Simon_57,Newell_94}.  A second,  not less important goal, is the evaluation and proposal of  strategies to improve the quality of those decisions. Accordingly,  in this paper we  revisit and generalize  a cooperative  problem-solving scenario introduced by Leavitt  in the  early 1950s \cite{Leavitt_51} (see also \cite{Bavelas_50}),  aiming at finding  the decision-making strategies that optimize the performance of the  group.

In Leavitt's  problem-solving  scenario, each of  $N$ subjects is dealt a deck of $N$ cards selected from a master set of $N+1$  cards. The setup is such that each card appears  on $N-1$ decks,  except for one card that appears on all $N$  decks. The group's task is  to find the common card in the shortest time possible \cite{Leavitt_51}. The original experiments  employed $N=5$ subjects who were placed on  cubicles that allowed the realization of a variety of communication patterns.  Actually, this setup was originally devised  and subsequently  widely used to   study   the  effects of the communication patterns on group performance \cite{Heise_51,Guetzkow_55,Shaw_54,Mulder_60} (see  \cite{Mason_12,Sebastian_17} for more recent contributions in this line).
 The subjects communicated by writing messages that could be passed through slots in the walls of the cubicles.
The task was considered finished when each agent indicated  the common card by throwing an appropriate switch.

The nature of  Leavitt's task contrasts with the problems usually posed to cooperative problem-solving  systems,  which  can also be solved by a single agent or by agents working independently of each other \cite{Clearwater_91,Kennedy_99,Lazer_07,Fontanari_15,Fontanari_18}. In fact,  cooperation among the  members of the group is mandatory to  find the  common card in the decks. In particular,  the typical  behavior of a subject in working toward a solution was to show all  its  cards  to the  members of the group with whom it was allowed to communicate, who then used that information to update their confidences about which card of their decks was the common card \cite{Leavitt_51}.  

Rather than focusing on small groups of human  subjects  as in the seminal works of the 1950s, here we consider extensive agent-based simulations aiming at offering a complete perspective on the group performance  for general $N$. In time, we assume that  the group's choice of the common card is determined by the majority rule with eventual ties  broken randomly, and that the group performance is measured by the failure probability, i.e., the probability the group chooses a wrong card. 

In addition, we make  Leavitt's task more difficult,  as well as more interesting from the perspective of decision making, by limiting to $C \leq N$ the number of cards the agents (i.e., our virtual subjects) can display  and by asking the group to make a decision at any time $\tau$. The important feature is that now  the agents must also decide which cards to show  to the other members of the group.  Here we consider two possibilities: the agents simply sample $C$  cards at random from their decks or they pick the sample they estimate is the most likely to contain the correct card. Of course, an agent's confidence that a particular card is the correct one is determined by  the number of times it observed that card in the decks of the  other agents. These two possibilities can be viewed as extremes of a  Gibbs probability distribution where the temperature  $T$, which must be  interpreted  simply as a noise parameter, measures  the strength of a random perturbation that prevents the agents to   rank  correctly the cards  according to their confidences.

Since our interest here is on the evaluation of decision-making strategies rather than on the effects  of communication patterns,  we assume that an agent can, in principle,   interact with any  other member of the group, so that   we select  the pairs of interacting agents at  random among the members of the group.
 Somewhat surprisingly, we find that the zero-temperature, noiseless  limit, in which the agents display only their highest confidence cards,   yields an extremely poor  performance for any time $\tau$. In addition,  we find that in the asymptotic time  limit  only the infinite temperature limit, in which the cards are displayed without regard to their confidences,
 guarantees that the correct card is found in all experiments. In  this case,  we find that the probability of failure decreases exponentially with increasing time  and  that the time constant $\tau_c$ scales with $N^{3/2}$ for $C<N$ and with $N$ for $C=N$. For a fixed time $\tau$, however, intermediate values of the temperature  yield a performance several orders of magnitude superior to the performances of the two temperature extremes. This surprising result reveals the importance of the agent's decision-making strategies  to  the optimization of   group performance.

The rest of this paper is organized as follows. In Section \ref{sec:model} we describe our generalization of Leavitt's task and the  rules that govern  the behavior of the agents in their attempts to  complete  the task.  
In  Section \ref{sec:res} we present and analyze the results of the agent-based simulations,  emphasizing the two extreme cases where the  displayed  cards  are drawn randomly without regard to their confidences ($T \to \infty$)  and where  they are chosen deterministically according to their confidence ranks ($T=0$).   
  Finally, in Section \ref{sec:disc}  we summarize our main findings and  compare them with the experimental results.


\section{Model}\label{sec:model}

We consider $N$ card  decks where  each deck contains  $N$ distinct cards. The cards of a deck  are selected from a  collection of $N+1$ distinct cards, so that  there is only one card  that is common to all decks and  any two decks have  exactly $N-1$ cards in common.
These decks are distributed to a group of $N$ agents (one deck for each agent)  and the task of the group  is to find the unique card that appears in all decks.  The agents must achieve that by exchanging information about their  card decks  and formulating a strategy to decide each of their $N$ cards is the common one. Next we describe the interaction and decision-making processes.

Each agent $i =1, \ldots, N$ assigns a confidence value to each card $\alpha_i =1, \ldots, N$ in its deck, denoted by the integer variable
$F_{i \alpha_i}^\tau = 0, 1, \ldots, \tau$, where $\tau$ stands for the number of interaction  rounds or, more simply, the time. This quantity measures the confidence of agent $i$ that  card $\alpha_i$ is the common card in the decks. The card labels have a subscript signaling the deck (or the agent) they belong to, since the decks do  not  have the same cards and there is no point for an agent to assign a confidence value  to a card that is not in its deck as that card surely is not the card common to all decks. 
 Hence, when agent $i$   is prompted to answer  which card is the common one at time $\tau$,  it  selects the card that maximizes the confidences  $F_{i \alpha_i}^\tau$. In case of a draw, a card is selected at random  among the tied cards. Initially all cards have the same confidence values, viz., $F_{i \alpha_i}^0 = 0$  for all $i$ and $\alpha_i$.  
The effect of the interaction between agents is then to change these confidences, as we describe next. 

We impose no restrictions on the communication between agents so, in principle, an agent  can interact with all its $N-1$ peers in the group. More pointedly, at time $\tau$ we choose randomly two distinct agents in the group.  The first agent - the observer -  assesses $C$ cards  of the second agent  - the observed -  and  updates its confidences by adding one unit of confidence to the cards they have in common. 
An interaction round is complete when this procedure is repeated $N$ times, so that $N$, not necessarily distinct, agents have updated their confidences. (We note that only the observers update their confidences.) Then the time counter is incremented by one unit, i.e, $\tau \to \tau +1$.

Since it is the observed agent that selects the $C$ cards that are assessed by the observer,  we need to set up strategies for selecting those cards. The  most natural strategy is simply  to sample $C$ cards without replacement from the deck of the observed agent.  Another natural strategy is to choose the $C$ cards with the highest confidences. As already  pointed out, these two strategies can be seen as extremes of the temperature of a Gibbs  distribution as follows. We recall that there are $\binom{N}{C}$  different samples of $C$ cards that can be drawn from a deck of $N$ cards. Because each agent  holds a different deck, we denote those samples by $\mathcal{C}_{k_j}$ with $k_j=1, \ldots,  \binom{N}{C}$, where $j$ is the label of the agent. Then we  associate an energy value to each  sample,
\begin{equation}\label{energy}
E \left ( \mathcal{C}_{k_j} \right ) = - \sum_{\alpha_j \in \mathcal{C}_{k_j}} F^\tau_{j \alpha_j} ,
\end{equation}
which is minimized by the sample containing  the cards with the highest confidence values. Here we assume that a sample $\mathcal{C}_{k_j}$ is selected with probability
\begin{equation}\label{prob}
\mathcal{P} \left  ( \mathcal{C}_{k_j} \right )  = \frac{1}{Z_j} \exp \left [ - \beta E \left ( \mathcal{C}_{k_j} \right ) \right ] 
\end{equation}
where $Z_j = \sum_{\mathcal{C}_{k_j}} \exp \left [ - \beta E \left ( \mathcal{C}_{k_j} \right ) \right ] $ is a normalization factor. 
The temperature $T = 1/\beta  \geq 0 $ is  interpreted as a measure of a noise that prevents the observed agent to select accurately the cards with the highest confidences. This parameter allows us to interpolate between the random sampling of cards ($T \to \infty$) and the selection of the cards with  the  highest confidences ($T= 0$).  

 Our introduction of the sample size 
$C=1,2, \ldots, N$ is justified when $N$ is large and  so it would take too long for the observer to scan and record all  cards in the hand of the observed agent. In any event, here we view it as an additional rule of the interaction between agents, i.e.,  that the observer can view only $C$ cards of the deck of the observed agent.

The decision of the group at time $\tau$ is obtained by asking each agent $i$ to designate its guess of the common card. As  pointed out before, this is done by picking the card $\alpha_i$ that maximizes the confidences $F_{i \alpha_i}^\tau$.  The group guess is then determined by the majority rule, i.e., the group picks  the card designated by the majority of the agents.  As usual, in case of a draw, the decision is made by selecting randomly one of the tied cards.  Here we measure the group performance by the failure probability $P_\tau$, which is given by the fraction of  independent experiments that failed to find the common card at time $\tau$. We use the failure probability, rather than the 
success probability  $1- P_\tau $, because it is more convenient for the purpose of quantifying its time asymptotic behavior in the case of efficient decision strategies.

In sum, the  group dynamics proceeds as follows. At $\tau =0$  we assign the card decks to the agents and each agent sets to zero the confidences of its cards. Next we choose randomly the observer and the observed agent. The observed agent evaluates the energy  of each of the $\binom{N}{C}$ samples of $C$ cards and picks a sample with a probability given by the Gibbs distribution (\ref{prob}). The observer then updates the confidences of the cards that appear in the  sample selected by the observed agent (i.e., the cards common to the sample and to the observer's deck).  The time $\tau$  increases by one unit only after this procedure is repeated $N$ times.

\section{Results}\label{sec:res}

 The initial setup of the $N$ card decks  is deterministic, i.e., all experiments begin with the same decks, as described in Section \ref{sec:model}. The experiments have two sources of stochasticity, viz., the choice of the pairs of interacting agents and the sample of $C$ cards scanned by the observer. In the case $C=N$, stochasticity is due only to the selection of the interacting agents. The results presented here represent  averages over $10^5$ independent runs.  Next we discuss separately from the general case  the two temperature extremes of the Gibbs distribution (\ref{prob}) as  their study  does not require the time-consuming  generation of the  $\binom{N}{C}$ cards samples.
 
\begin{figure} 
\centering  
 \includegraphics[width=0.48\textwidth]{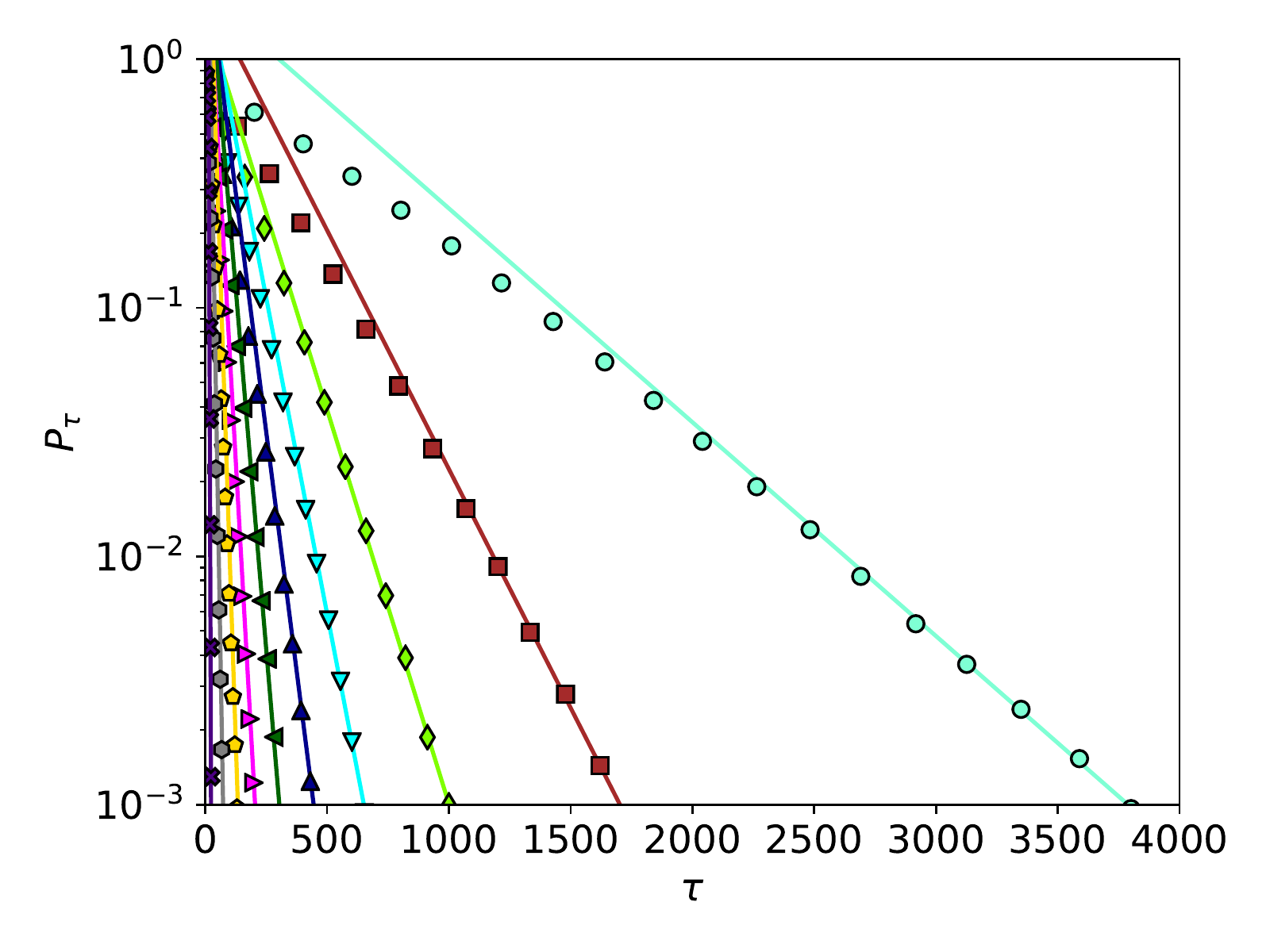}  
\caption{Failure probability $P_\tau $ at time $\tau$,  for  $N=10$  and (right to left) $C=1,2,\ldots, 10$ in the infinite temperature limit.
The lines are the fittings $P_\tau  = a_{N,C} \exp \left ( -\tau/\tau_c \right )$ where $ a_{N,C}$ and $\tau_c= \tau_c \left (N,C\right )$ are fitting parameters.
 }  
\label{fig:1}  
\end{figure}

\subsection{Infinite temperature limit}

Here we consider the case where the observed agent  randomly chooses a  sample of $C$ cards from its deck without regard to  their confidences, and offers them to the observer for  inspection. This corresponds to the limit $T=1/\beta \to \infty$ of the Gibbs distribution (\ref{prob}). This information is then used to update the observer's cards confidences as discussed before. Figure \ref{fig:1}  shows the dependence of the probability  of failure $P_\tau$ on the time $\tau$ for fixed $N$ in a semi-log plot.  Except for  small $\tau$,  this probability exhibits a neat
exponential decay  characterized by the time constant  $\tau_c = \tau_c \left (N,C\right )$, which  measures the typical time the group requires to  complete successfully  the task. This figure   shows that, as expected, $\tau_c$ increases as  the sample size  $C$ decreases and, more importantly, that given enough time  the group is always guaranteed to find the common card, i.e., $P_\tau \to 0$ for $\tau \to \infty$.

\begin{figure} 
\centering  
 \includegraphics[width=0.48\textwidth]{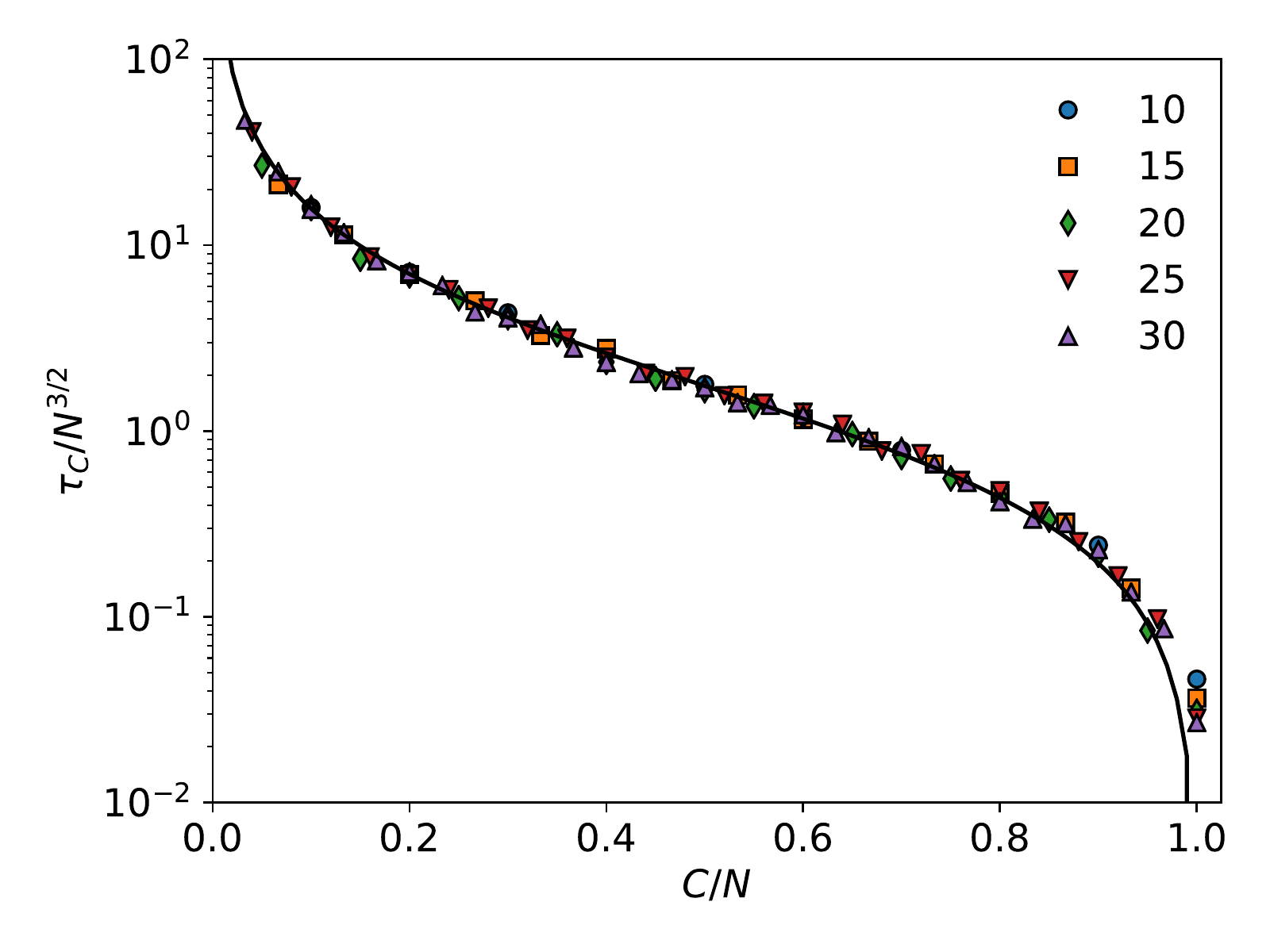}  
\caption{Semi-logarithm plot of the time constant $\tau_c$ as function of the ratio $C/N$ for  the infinite temperature limit and
$N=10, 15, 20, 25, 30$ as indicated.
The curve is the fitting $\tau_c = 1.75 \left  [ 1/\left (C/N \right ) - 1  \right ]$.
 }  
\label{fig:2}  
\end{figure}

\begin{figure} 
\centering  
 \includegraphics[width=0.48\textwidth]{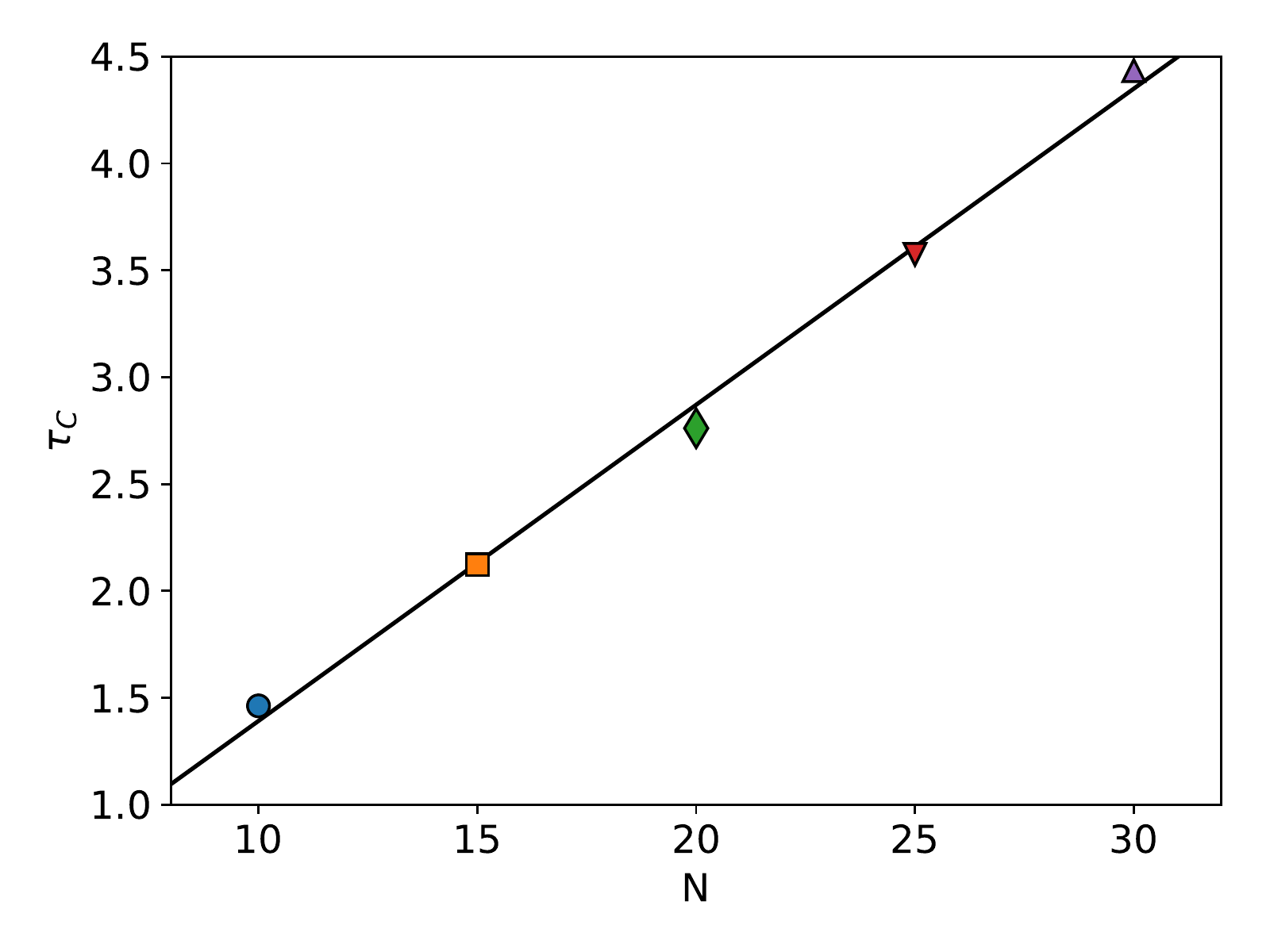}  
\caption{Dependence of the time constant $\tau_c$ on the group or deck size $N$  for $C=N$ in the infinite temperature limit. The symbols convention is the same as
for Fig.\ \ref{fig:2}. The straight line is the
 fitting $\tau_c = -0.09 + 0.15 N$.
 }  
\label{fig:3}  
\end{figure}

The dependence of the time constant $\tau_c$ on the parameters of the model is exhibited  more clearly  in  Fig.\ \ref{fig:2}, which shows that the data for different $N$ are described very well by the scaling form
\begin{equation}\label{tc}
\tau_c = N^{3/2} f \left ( C/N \right) ,
\end{equation}
with the scaling function $f(x) = 1.75 \left  (1/x - 1  \right ) $. In this figure we use a semi-logarithm plot  to highlight the goodness of  fit in the region of $C/N$ close to 1.
 We note that in the case  $C=N$,  $\tau_c$ increases linearly with increasing  $N$ as shown in Fig.\ \ref{fig:3}. This result is consistent with eq.\ (\ref{tc}) because $f(1)=0$  and so this equation  does not determine the scaling of $\tau_c$ with $N$ in the case $C=N$. Interestingly, the  uncertainty introduced by reducing the amount of information the agents can transmit to each other changes the scale of the  time constant with the group or deck size.

\subsection{Zero temperature limit}

Since in each interaction the observer  inspects only  $C$  cards of the deck of the observed agent, a natural   strategy to improve the  performance of the group is to  guarantee that the latter displays the sample of cards that  is the  most likely to contain the  correct  card.   An apparently sensible  strategy  is to choose the $C$ cards with the highest confidence values.  This can be achieved by considering  the zero temperature  $T = 1/\beta \to 0$ limit of the Gibbs distribution (\ref{prob}), but it is much more efficient simply to rank the  card confidences $F_{j \alpha_j}^\tau$;  $\alpha_j =1, \ldots, N$  of the observed agent $j$  and select the top $C$ cards. Eventual draws between cards in and out this selection are broken randomly, as usual.

\begin{figure} 
\centering  
 \includegraphics[width=0.48\textwidth]{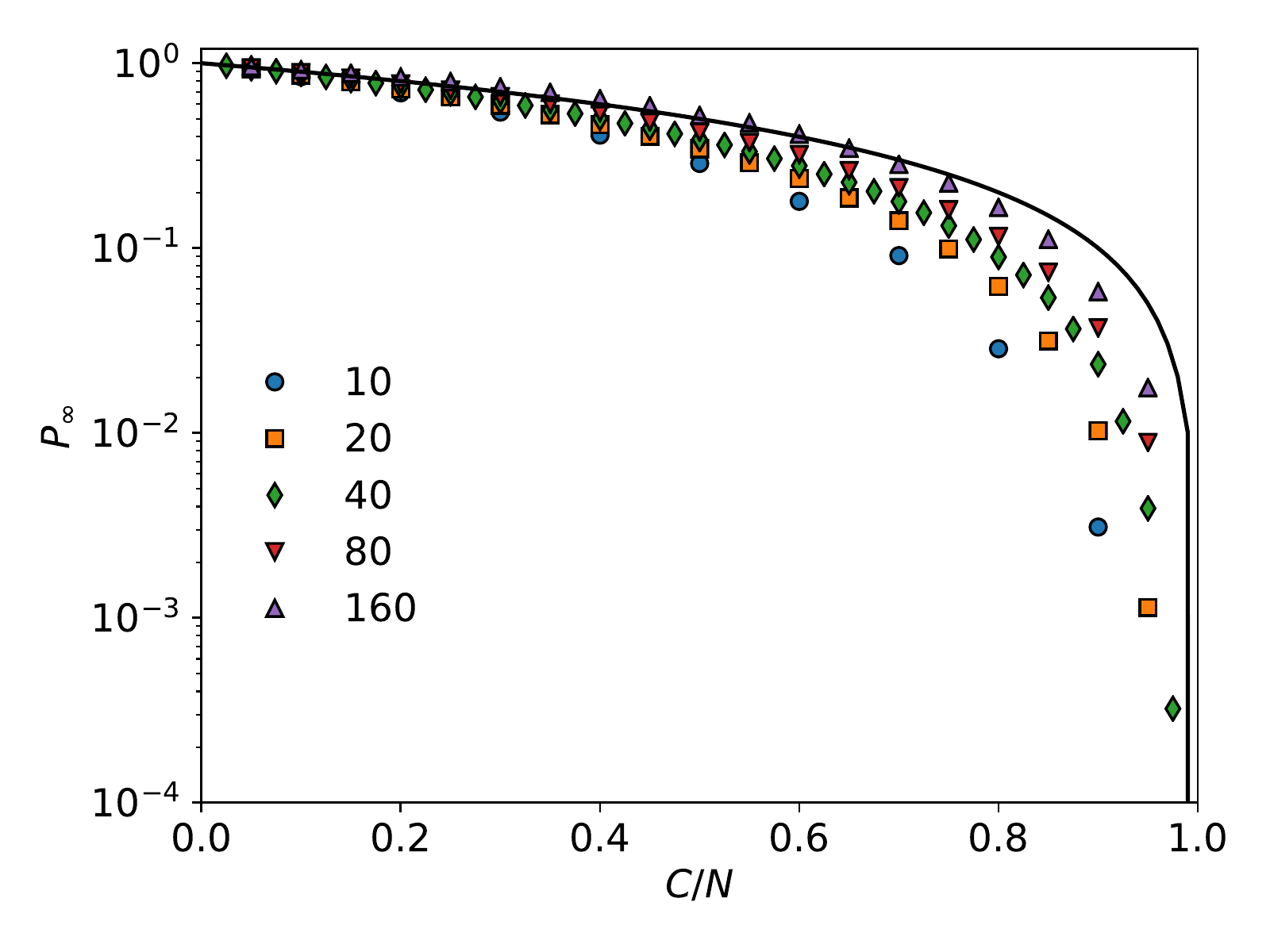}  
\caption{Asymptotic probability of failure  $P_\infty $  as function of the ratio $C/N$ for  the  zero temperature limit and
$N=10, 20, 40, 80, 160$ as indicated.
The solid curve is the function $P_\infty = 1 - C/N$ obtained by  extrapolating the finite $N$ results to $N \to \infty$.
 }  
\label{fig:4}  
\end{figure}

We find that  displaying  the $C$ cards with the highest confidences   is actually disastrous since the probability of failure rapidly tends to  a nonzero limiting value and becomes insensitive to increasing $\tau$. (This point will become clearer in the study of the finite temperature case.) Accordingly, in  Fig.\ \ref{fig:4} we show the asymptotic failure probability $P_\infty$ for several deck sizes $N$. Extrapolation of these data  to $N \to \infty$  yields the simple result $P_\infty = 1 - C/N$, which offers a clue to the failure of this noiseless strategy.
In fact,  there is the sporting chance $\left (N-C \right )/N = 1 - C/N$  that the common card is not chosen by an agent in the first round when the confidences are equiprobable and the results indicate that  for large $N$ this agent can  strongly influence the other agents  towards a wrong decision.

\subsection{Finite temperature}

Here we consider the more general scheme that encompasses both  strategies presented before, in which the $C$ cards shown to the observer are selected according to the Gibbs distribution (\ref{prob}) using a finite and  nonzero temperature.  The implementation of this case is more demanding computationally as we need to  calculate the  probabilities associated to all  $\binom{N}{C}$  samples  of $C$ cards that can be   drawn from the deck of the observed agent.

\begin{figure} 
\centering  
 \includegraphics[width=0.48\textwidth]{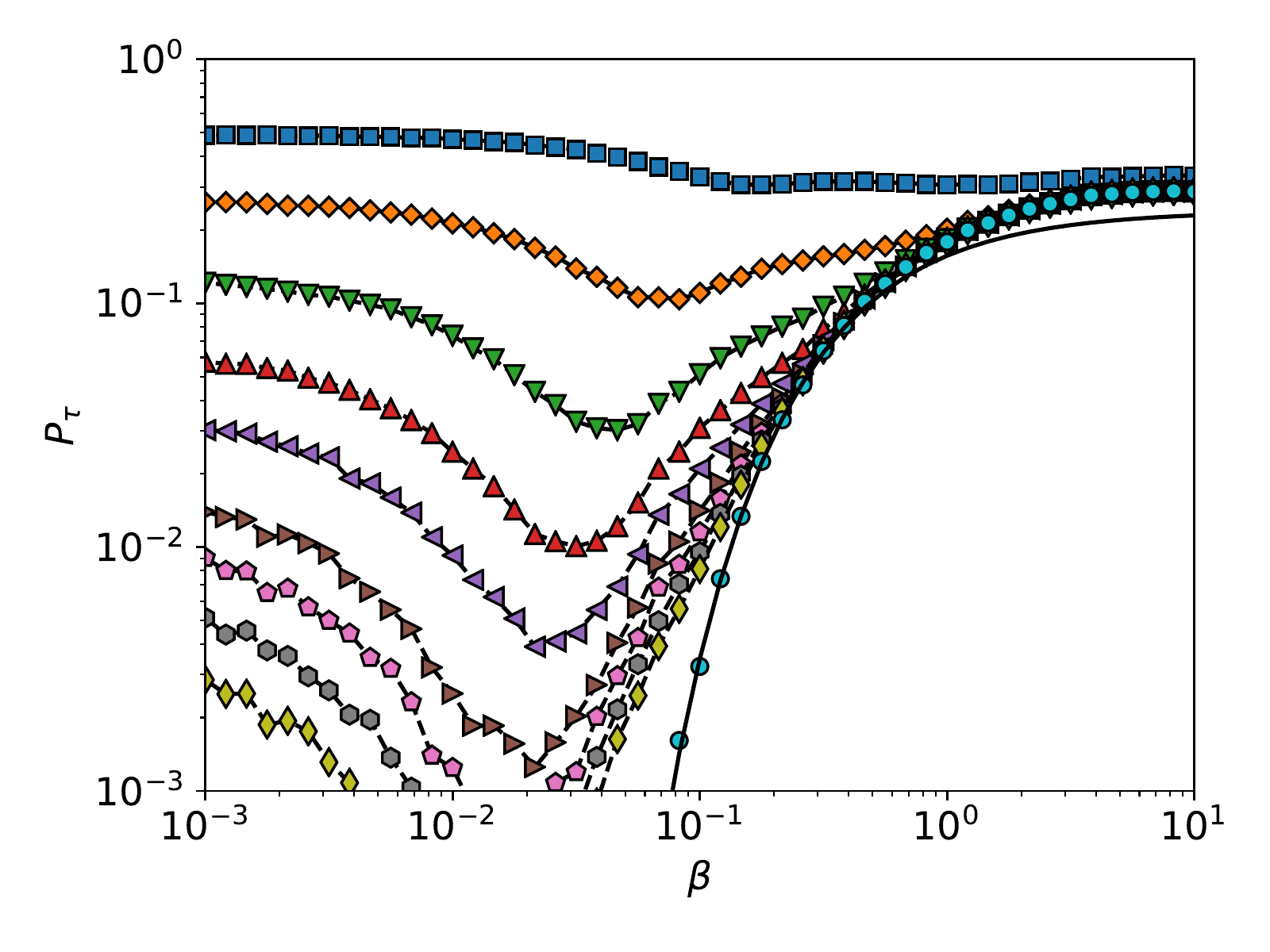}  
\caption{Probability of failure  $P_\tau $  as function of the inverse temperature $\beta$  for  $N=10$  and $C=5$ for (top to bottom)
$\tau = 43, 91, 144, 190, 229, 275, 301, 331, 363$ and $10^4$.   In the limit 
$\beta \to \infty$,  we find   $P_{\infty} = 0.287$.
The solid  curve is the  fitting of the data for $\tau = 10^4$,  $P_{10^4}  =  a \exp \left ( -b/\beta \right )$ where the  fitting parameters are  $ a = 0.251$ and $b=0.428$.
 }  
\label{fig:5}  
\end{figure}

Figure \ref{fig:5}, which shows the probability of failure $P_\tau$ as function of the  inverse temperature  $\beta$ at different  times $\tau$, reveals a few surprises.  First, except for very short times  the noiseless strategy exhibits the worst performance,  as mentioned before.
Second, for fixed $\tau$ there is a nonzero value of $\beta$ that minimizes the failure probability. For instance, in Fig.\ \ref{fig:5} the failure probability for $\tau = 363$ at $\beta = 0.01$  is several orders of magnitude lower than at $\beta =0$. Third  and  most surprisingly, the asymptotic failure probability  vanishes for $\beta =0$ only. More pointedly,  we find $P_\infty \asymp \exp \left ( - b/\beta \right )$ where $b$ is a parameter that depends on $N$ and $C$. This means that the value of $\beta$ that minimizes $P_\tau$ tends to zero as $\tau$ increases, a trend that is already visible in the figure. We have verified that these conclusions hold true for all values of the parameters $N$ and $C<N$.

 \section{Discussion}\label{sec:disc}
 
Understanding the  influences on  group performance is a central issue in collective intelligence \cite{Page_07}.  Most  theoretical and experimental efforts have focused on the influence of the composition of the group, since this factor can be more easily observed and controlled \cite{Levine_93,Woolley_10}. Here we focus instead on the strategies adopted  by   individuals to select and transmit relevant information to the other members of the group.  

More pointedly,
we  revisit Leavitt's cooperative problem-solving  scenario in which each member of a group of size $N$  receives a deck with $N$ cards and its task is to find  the  card that is common to all decks in the shortest time \cite{Leavitt_51,Bavelas_50}. The decks are contrived so as to maximize the difficulty of the task, e.g.,  any two decks have $N-1$ cards  in  common and the cards come in  $N+1$ types.
Whereas the experiments carried out in the 1950s  employed   $N=5$ subjects in a variety of communication patterns (i.e., who can communicate with whom),  here we use agent-based simulations to study the group performance for general $N$ in a fully connected communication net. We ground our study on the persuasive assumption that an agent's confidence that a particular card is the correct one is determined by  the number of times it observed that card in the decks of the  other agents. In fact, this strategy of increasing the  card confidences  following  the agent interactions  is akin to the scheme used to model lexicon acquisition  in a cross-situational learning scenario \cite{Tilles_12,Reisenauer_13}. In addition, we assume that the group decision is determined by applying the majority rule to the decisions of the group members. 

In the case the agents can assess all cards of the decks of the  other members, we find that the  scale of time   to complete successfully  the task grows linearly with the number of agents or decks, i.e., $\tau_c \propto N$ (see Fig.\ \ref{fig:3}). More pointedly,  we find  that the probability of failure decreases exponentially with increasing time, $P_\tau \propto \exp \left  (-\tau/\tau_c \right )$. This exponential decay is a consequence of the fact that the interacting agents are chosen randomly, leading to the possibility of uninformative repeated interactions or of one or more agents never being selected for interaction in a finite time $\tau$.  

The group task is made considerably more  difficult by limiting  to $C<N$ the number of cards that the agents can show to each other. In fact, in the  case the $C$ cards are sampled randomly without regard to their confidences, the  time constant becomes  much  greater than in the unrestricted case $C=N$, namely, $\tau_c \propto N^{3/2}$ (see Fig.\ \ref{fig:2}). An interesting  consequence of  restricting the number of  cards  displayed is the need to  specify a strategy  to select those  cards.  A natural line that we pursue here is 
to link  the  chances of display of a card to its confidence. We find that the  exhibition of the cards with the highest confidences  yields a very poor performance since  it maximizes the failure probability for large $\tau$ (see Fig.\  \ref{fig:5}). The importance of the confidences on the selection of the displayed cards can be tuned by introducing  the Gibbs probability distribution  (\ref{prob}), where the  temperature $T= 1/\beta$ measures the strength of a  noise that prevents the correct ranking of the cards.  Most interestingly, for fixed $\tau$ we find that  the failure probability is minimized for a finite, nonzero temperature (see Fig.\ \ref{fig:5}). However, in the asymptotic limit $\tau \to \infty$ it vanishes only in the infinite temperature limit, i.e., in the case the cards confidences do not affect their chances of being   selected for display.

Although the two temperature extremes of the Gibbs distribution (\ref{prob}) are sensible strategies the subjects could adopt in a real-world situation, the finite   temperature strategies are clearly too complex to be used in practice. Of course,  our intention of introducing the noise parameter or temperature  was simply to add an element  that could prevent the correct ranking of the cards and, as physicists, the Gibbs distribution came along naturally.  We expect, however,  that a more psychologically realistic implementation of this noise would produce similar results.

Despite the simplicity of Leavitt's task, it does not yield to an analytical approach even in the infinite temperature limit, where the card confidences play no role on the  decisions of the agents.  In fact, a natural procedure to tackle this problem analytically is to write recursion equations for the probabilities that the  card confidences take on the (integer) values $0, \ldots, \tau$ at time $\tau$  \cite{Feller_68}. The difficulty is that the card confidences are not independent random variables  and so we need to work with the  joint probability distribution of
the $N$ card confidences, which is impractical even for small $N$. Hence  our Monte Carlo approach is probably  the simplest, if not the only,  way to study  Leavitt's task and its   generalizations proposed here. Nevertheless, our use of scaling forms to analyze the simulation results, which is perhaps the  main contribution of physics to the so-called science of complexity  \cite{West_17},  offered  a thorough characterization of the problem. 

\begin{table}[]
\caption{Asymptotic mean number of runs with at least one error    in the case of $N=5$ agents fixed at the corners of a pentagon. The experimental result is  $\eta =  0.23$ \cite{Leavitt_51}. }
\centering
\begin{tabular}{cccccc}
\hline
 & $\beta =0$  & $\beta =0.1$ & $\beta =0.3$ & $\beta =0.5$ & $\beta \to \infty$ \\ \hline
C=1 & 0.998     &  0.251     &  0.503   & 0.527      & 0.601            \\
C=2 & 0.997     &  0.058     &  0.201   & 0.267	& 0.387            \\
C=3 & 0.996     &  0.004     &  0.149   &	0.302   	& 0.549            \\
C=4 & 0.997     &  0.001     &  0.092   & 	0.198	& 0.565            \\
C=5 & 0.997     &  0.999     &  0.997   &	0.996	& 0.997        \\
\hline   
\end{tabular}
\label{table:1}
\end{table}
   
To conclude, a comparison of our results with the seminal experiments  of Leavitt \cite{Leavitt_51} is in order. As pointed out, the  experimental studies considered $N=5$ subjects, who were positioned at the corners of  geometric patterns (e.g., a pentagon).  In addition, there were no  restrictions on the number of cards that could be displayed (i.e., $C$ was not fixed and could vary in time as well as from subject to subject). More importantly, Leavitt did not focus on the group decision as we did here, but on the individual subjects' decisions. 
An unambiguous outcome of   Leavitt's experiments is the mean number of experiments with at least one error, which we denote by $\eta$. An error occurs when a subject  selects a wrong card. In the case of a pentagon,  Leavitt found $\eta =  0.23$. Our simulation results for  this scenario are exhibited in Table \ref{table:1} where we varied the sample size $C$ and the inverse  temperature $\beta$. The results indicate that for any value of $C<5$ it is possible to find a value of $\beta$ that reproduces the experimental result, which, however, cannot be reproduced by any of the temperature extremes. This finding  highlights the importance of (moderate) noise to model the human mental processes of decision making.

The simplicity of Leavitt's task and the prospect of modeling the decision-making process of the individuals is an enthralling research avenue that calls for more experimental results using, perhaps, digital platforms that are not limited by the physical constraints of the experiments carried out in the 1950s. We hope this contribution will motivate further research on this direction.

\acknowledgments
The research of JFF was  supported in part 
 by Grant No.\  2017/23288-0, Fun\-da\-\c{c}\~ao de Amparo \`a Pesquisa do Estado de S\~ao Paulo 
(FAPESP) and  by Grant No.\ 305058/2017-7, Conselho Nacional de Desenvolvimento 
Cient\'{\i}\-fi\-co e Tecnol\'ogico (CNPq).
SMR  was supported by grant  	15/17277-0, Fun\-da\-\c{c}\~ao de Amparo \`a Pesquisa do Estado de S\~ao Paulo 
(FAPESP).

\end{document}